\documentclass[12pt]{iopart}
\usepackage[english]{babel}
     \newcommand{\w}{\wedge}
     \newcommand{\wt}{\widetilde}
\begin{document}

\title[Equivalence of Hunter - Saxton Equation and
Euler - Poisson Equation]
{Contact Equivalence of the Generalized Hunter - Saxton Equation and
the Euler - Poisson Equation}

\author{Oleg I Morozov}

\address{Department of Mathematics, Moscow State Technical University
of Civil Aviation,\\20 Kronshtadtsky Blvd, Moscow 125993, Russia}
\ead{oim@foxcub.org}

\begin{abstract}
We present a contact transformation of the generalized Hunter--Saxton
equa\-ti\-on to the Euler--Poisson equation with spe\-ci\-al values of the
Ovsiannikov in\-va\-ri\-ants. We also find the general solution for the
generalized Hunter--Saxton equa\-ti\-on.
\end{abstract}

\ams{58H05, 58J70, 35A30}

\nosections

The generalized Hunter--Saxton equation
\begin{equation}
u_{tx} = u\,u_{xx}+\kappa\,u_x^2
\label{HS}
\end{equation}
has a number of applications in the nonlinear instability theory of a director
field of a liquid crystal, \cite{HunterSaxton}, in geometry of Einstein--Weil
spaces, \cite{Tod,Dryuma}, in constructing partially invariant solutions for
the Euler equations of an ideal fluid, \cite{Golovin}, and has been a subject
of many recent studies. In the case $\kappa = \case{1}{2}$ the general
solution, \cite{HunterSaxton}, the tri-Hamiltonian formulation,
\cite{OlverRosenau}, the pseudo-spherical formulation and the quadratic
pseudo-potentials, \cite{Reyes}, have been found. The conjecture of
linearizability of equa\-ti\-on (\ref{HS}) in the case $\kappa = - 1$ has been
made in \cite{Golovin}. In \cite{Pavlov}, a formula for the general solution
of (\ref{HS}) has been proposed. This formula uses a nonlocal change of
variables.

In this paper, we prove that equation (\ref{HS}) is equivalent under a contact
trans\-for\-ma\-ti\-on to the Euler--Poisson equation,
\cite[\S~9.6]{Ovsiannikov},
\begin{equation}
u_{tx} = {{1}\over{\kappa\,(t+x)}}\,u_t
+ {{2\,(1-\kappa)}\over{\kappa\,(t+x)}}\,u_x
- {{2\,(1-\kappa)}\over{(\kappa\,(t+x))^{2}}}\,u,
\label{EP}
\end{equation}
and find the general solution of (\ref{HS}) in terms of local variables.

In \cite{Morozov2004_2}, \'E. Cartan's method of equivalence,
\cite{Cartan2}--\cite{Cartan5}, \cite{Gardner,Olver}, in its form of the moving coframe method,
\cite{FelsOlver,Morozov2002,Morozov2004_1}, was used to find the
Maurer--Cartan forms for the pseudo-group of contact symmetries of
equa\-ti\-on (\ref{EP}). The structure equa\-ti\-ons for the symmetry
pseudo-group have the form
\begin{eqnarray}
\rmd\theta_0 &=
\eta_1\w\theta_0
+\xi^1\w\theta_1
+\xi^2\w\theta_2,
\nonumber \\
\rmd\theta_1 &=
\eta_2\w\theta_1
-2\,(1-\kappa)\,\theta_0\w\xi^2
+\xi^1\w\sigma_{11},
\nonumber \\
\rmd\theta_2 &=
(2\,\eta_1-\eta_2)\w\theta_2
-\theta_0\w\xi^1
+\xi^2\w\sigma_{22},
\nonumber \\
\rmd\xi^1 &=
(\eta_1-\eta_2)\w\xi^1,
\nonumber \\
\rmd\xi^2 &=
(\eta_2 -\eta_1)\w\xi^2,
\label{SE}
\\
\rmd\sigma_{11} &=
(2\,\eta_2-\eta_1)\w\sigma_{11}
+\eta_3\w\xi^1
+3\,(2\,\kappa-1)\,\theta_1\w\xi^2,
\nonumber \\
\rmd\sigma_{22} &=
(3\,\eta_1-2\,\eta_2)\w\sigma_{22}
+\eta_4\w\xi^2,
\nonumber \\
\rmd\eta_1 &=
(2\,\kappa-1)\,\xi^1\w\xi^2,
\nonumber \\
\rmd\eta_2 &=
(1-4\,\kappa)\,\xi^1\w\xi^2,
\nonumber \\
\rmd\eta_3 &=
\pi_1\w\xi^1
- (2\,\eta_1-3\,\eta_2)\w\eta_3
+4\,(3\,\kappa-1)\,\xi^2\w\sigma_{11},
\nonumber \\
\rmd\eta_4 &=
\pi_2\w\xi^2
+(4\,\eta_1-3\,\eta_2)\w\eta_4
+2\,(3-\kappa)\,\xi^1\w\sigma_{22},
\nonumber
\end{eqnarray}
where 
$\theta_0$, $\theta_1$, $\theta_2$, $\xi^1$, $\xi^2$, $\sigma_{11}$,
$\sigma_{22}$, $\eta_1$, ... , $\eta_4$ are the Maurer--Cartan forms,
while $\pi_1$ and $\pi_2$ are 
prolongation forms. We have
$\theta_0 = a\,(\rmd u-u_t\,\rmd t-u_x\,\rmd x)$,
$\theta_1 = a\,b^{-1}(\rmd u_t - u_{tt}\rmd t -R_2\,\rmd x)
+ 2\,(\kappa-1)\,(\kappa\,b\,(t+x))^{-1}\theta_0$,
$\theta_2 =a\,b\,\kappa\,(t+x)^2\,(\rmd u_x - R_2\,\rmd t- u_{xx}\,\rmd x)
+ b\,(t+x)\,\theta_0$,
$\xi^1 = b\,\rmd t$, and $\xi^2 = b^{-1} \kappa^{-1} (t+x)^{-2} \rmd x$,
where $R_2$ is the right-hand side of equation (\ref{EP}),
while $a$ and  $b$ are arbitrary non-zero constants.
The forms $\sigma_{11}$, ... , $\pi_2$ are too long to be written out
in full here.
We write equation (\ref{HS}) and its Maurer--Cartan forms in tilded
variables, then similar computations give
$\wt{\theta}_0 = \wt{a}\,(\rmd\wt{u}-\wt{u}_{\wt{t}}\,\rmd\wt{t}
-\wt{u}_{\wt{x}}\,\rmd\wt{x})$,
$\wt{\theta}_1 = \wt{a}\,\wt{b}^{-1}
(\rmd\wt{u}_{\wt{t}} - \wt{u}_{\wt{t}\wt{t}}\,\rmd\wt{t} -
{\wt{R}}_1\,\rmd\wt{x})
- \wt{b}^{-2} \wt{u}\,\wt{u}_{\wt{x}\wt{x}}\,\, \wt{\theta}_2
- (2\,\kappa-1)\,\wt{b}\,\wt{u}_{\wt{x}}\,\,\wt{\theta}_0$,
$\wt{\theta}_2 = \wt{a}\,\wt{b}^{-1}(\wt{u}_{\wt{x}\wt{x}})^{-1}\,
(\rmd\wt{u}_{\wt{x}}-{\wt{R}}_1
\,\rmd\wt{t} - \wt{u}_{\wt{x}\wt{x}}\,\rmd\wt{x})$,
$\wt{\xi}^1 = \wt{b}\,\rmd\wt{t}$,
and
$\wt{\xi}^2 = \wt{b}^{-1}\,(\rmd\wt{u}_{\wt{x}}
- \kappa\,(\wt{u}_{\wt{x}})^2\,\rmd\wt{t})$,
where ${\wt{R}}_1$ is the right-hand side of equation (\ref{HS}) written in the tilded vatiables,
while ${\wt{a}}$ and  ${\wt{b}}$ are arbitrary non-zero constants.
The forms $\wt{\sigma}_{11}$, ... , $\wt{\pi}_2$ are too long to be written out
in full. 
The structure equations for (\ref{HS}) differ from (\ref{SE}) only in
replacing $\theta_0$, ... , $\pi_2$ by their tilded counterparts.
Therefore, results of Cartan's method (see, e.g.,
\cite[th~15.12]{Olver}) yield the contact equivalence of equations (\ref{HS})
and (\ref{EP}). Since the Maurer--Cartan forms for both symmetry groups
are known, the equivalence transformation $\Psi : (t,x,u,u_t,u_x) \mapsto
(\wt{t},\wt{x},\wt{u},\wt{u}_{\wt{t}},\wt{u}_{\wt{x}})$
can be found from the requirements $\Psi^{*}\wt{\theta}_0 = \theta_0$,
$\Psi^{*}\wt{\theta}_1 = \theta_1$, $\Psi^{*}\wt{\theta}_2 = \theta_2$,
$\Psi^{*}\wt{\xi}^1 = \xi^1$, and $\Psi^{*}\wt{\xi}^2 = \xi^2$:

\vskip 5 pt

{\bf Theorem.}
{\it
The contact transformation $\Psi$
\begin{eqnarray}
\wt{u} &=(t+x)^{-\frac{1}{\kappa}}\left(\kappa\,(t+x)\,u_x+(\kappa-1)\,u\right),
\nonumber\\
\wt{t} &= \kappa^{-1}\,t,
\nonumber\\
\wt{x} &= -(t+x)^{\frac{\kappa-1}{\kappa}}\left(\kappa\,(t+x)\,u_x- u\right),
\nonumber\\
\wt{u}_{\wt{t}} &= \kappa^2\,(t+x)^{-\frac{1}{\kappa}}\left(u_t-u_x \right),
\nonumber\\
\wt{u}_{\wt{x}} &= - (t+x)^{-1}
\nonumber
\end{eqnarray}
\noindent
takes the Euler--Poisson equation (\ref{EP}) to the generalized Hunter--Saxton
equa\-ti\-on (\ref{HS}) (writ\-ten in the tilded variables).
}

\vskip 5 pt

{\bf Remark.}
The equivalence transformation $\Psi$ is not uniquely determined: for any
$\Phi$ and $\Upsilon$ from (isomorphic) infinite-dimensional pseudo-groups of
contact symmetries of equations (\ref{HS}) and (\ref{EP}), respectively, the
transformation $\Phi \circ \Psi \circ \Upsilon$ is also an equivalence
transformation.

\vskip 5 pt

Equation (\ref{EP}) belongs to the class of linear hyperbolic equations
$u_{tx} = T(t,x)\,u_t + X(t,x)\,u_x + U(t,x)\,u$ and has important features:
it has an intermediate integral, and its general solution can be found in quadratures. 
To prove  this, we compute for equation (\ref{EP}) the Ovsiannikov invariants, \cite[\S~9.3]{Ovsiannikov}, $P=K\,H^{-1}$ and 
$Q =\left( \ln \vert H \vert \right)_{tx} H^{-1}$, where $H = -T_t+T\,X+U$ and 
$K = -X_x+T\,X+U$ are the Laplace semi-invariants. We have $P = 2\,(1-\kappa)$ and 
$Q = 2\,\kappa$, therefore $P + Q = 2$, and the Laplace $t$-transformation, \cite[\S~9.3]{Ovsiannikov}, takes equation (\ref{EP}) to a factorizable linear hyperbolic equation. Namely, we consider the system
\begin{eqnarray}
v &= u_x - (\kappa\,(t+x))^{-1} u,
\label{B1}
\\
v_t &= 2\,(1-\kappa)\,(\kappa\,(t+x))^{-1} v + \kappa^{-1}\,(t+x)^{-2} u.
\label{B2}
\end{eqnarray}
Substituting (\ref{B1}) into (\ref{B2}) yields equation (\ref{EP}), while
expressing $u$ from (\ref{B2}) and substituting it into (\ref{B1}) gives the
equation
\begin{equation}
v_{tx} =
  {{1- 2\,\kappa}\over{\kappa\,(t+x)}} \, v_t
+ {{2\,(\kappa-1)}\over{\kappa\,(t+x)}} \, v_x
- {{(2\,\kappa-1)\,(\kappa-2)}\over{(\kappa\,(t+x))^{2}}} v
\label{EP_trans}
\end{equation}
with the trivial Laplace semi-invariant $H$. Hence, the substitution
\begin{equation}
w = v_x + (2\,\kappa-1)\,(\kappa\,(t+x))^{-1} v
\label{B3}
\end{equation}
takes equation (\ref{EP_trans}) into the equation
\begin{equation}
w_t = - 2\,(\kappa-1)\,(\kappa\,(t+x))^{-1} w.
\label{B4}
\end{equation}
Integrating (\ref{B4}) and (\ref{B3}), we have the general solution for
equation (\ref{EP_trans}):
\[
v = (t+x)^{\frac{1-2\kappa}{\kappa}}\,\left(
S(t) + \int R(x)\,(t+x)^{\frac{1}{\kappa}}\,\rmd x
\right),
\]
where $S(t)$ and $R(x)$ are arbitrary smooth functions of their arguments.
Then equation (\ref{B2}) gives the general solution for equation (\ref{EP}):
\[\fl
u{=}
(t{+}x)^{\frac{1}{\kappa}}\,\left(
\kappa\,S^{\prime}(t){+}\int R(x)\,(t{+}x)^{\frac{1{-}\kappa}{\kappa}}\,\rmd x
\right){-}(t{+}x)^{\frac{1{-}\kappa}{\kappa}}\,
\left(
S(t){+}\int R(x)\,(t{+}x)^{\frac{1}{\kappa}}\,\rmd x
\right).
\]
This formula together with the contact transformation of the theorem gives
the general solution for the generalized Hunter--Saxton equation (\ref{HS})
in a parametric form:
\begin{eqnarray}
\wt{u} &= \kappa^2\,S^{\prime}(t)
+\kappa\,\int R(x)\,(t+x)^{\frac{1-\kappa}{\kappa}}\,\rmd x,
\nonumber \\
\wt{t} &= \kappa^{-1}\,t,
\nonumber \\
\wt{x} &= -\kappa\,
\left( S(t)
+\int R(x)\,(t+x)^{\frac{1}{\kappa}}\,\rmd x\right).
\nonumber
\end{eqnarray}

Hence,  we obtain the general solution of equation (\ref{HS}) without employing nonlocal transformations.


\section*{References}

\begin{thebibliography}{99}
\bibitem{HunterSaxton} Hunter J K and Saxton R 1991
  Dynamics of director fields
   {\it SIAM J. Appl. Math.} {\bf 51} 1498 - 521
\bibitem{Tod} Tod K P 2000
  Einstein--Weil spaces and third order differential equations
   {\it J. Math. Phys.} {\bf 41} 5572 - 81
\bibitem{Dryuma} Dryuma V 2001 On the Riemann and Einstein--Weil Geometry
   in Theory of the Second Order Ordinary Differential Equations
   {\it  Preprint} math.DG/0104278
\bibitem{Golovin} Golovin S V  2004
  Group Foliation of Euler Equations in Nonstationary Rotationally
  Symmetrical Case
   {\it Proc. Inst. Math. NAS of Ukraine}
   {\bf 50} Part 1, 110 - 7
\bibitem{OlverRosenau} Olver P J and Rosenau Ph  1996
  Tri-Hamiltonian duality between solitons and so\-li\-ta\-ry wave so\-lu\-ti\-ons having
  compact support
   {\it Phys Rev E} {\bf 53} 1900 - 6
\bibitem{Reyes} Reyes E G 2002
  The soliton content of the Camassa--Holm and Hunter--Saxton Equations
   {\it Proc. Inst. Math. NAS of Ukraine}
   {\bf 43} Part 1,  201 - 8
\bibitem{Pavlov} Pavlov M V 2001  The Calogero equation and Liouville type
   equations {\it  Preprint} nlin.SI/0101034
\bibitem{Ovsiannikov} Ovsiannikov L V 1982 {\it Group Analysis of
   Differential Equations} (New York: Academic Press)
\bibitem{Morozov2004_2} Morozov O I 2004 Contact Equivalence Problem for Linear
   Hyperbolic Equations  {\it Preprint}  math-ph/0406004
\bibitem{Cartan2} Cartan \'E 1953 {\it
        Les sous-groupes des groupes continus de transformations //
        {\OE}uvres Compl{\`e}tes}, Part II,  {\bf  2}
        (Paris: Gauthier - Villars)
        719--856
\bibitem{Cartan4} Cartan \'E 1953 {\it
        La structure des groupes infinis. //
        {\OE}uvres Compl{\`e}tes}, Part II,  {\bf  2}
        (Paris: Gauthier - Villars)
        1335--84
\bibitem{Cartan5} Cartan \'E 1953 {\it
        Les probl\`emes d'\'equivalence. //
        {\OE}uvres Compl{\`e}tes}, Part II,  {\bf  2}
        (Paris: Gauthier - Villars)
        1311--1334
\bibitem{Gardner} Gardner R B 1989 {\it The method of equivalence and its
   applications} (Philadelphia: SIAM)
\bibitem{Olver} Olver P J 1995 {\it Equivalence, Invariants, and Symmetry}
   (Cambridge: Cambridge University Press)
\bibitem{FelsOlver} Fels M, Olver P J 1998
  Moving coframes I. A practical algorithm
   {\it Acta Appl. Math.} {\bf 51} 161--213
\bibitem{Morozov2002} Morozov O I 2002
  Moving Coframes and Symmetries of Differential Equations
   {\it J. Phys. A: Math. Gen.} {\bf 35} 2965 -- 77
\bibitem{Morozov2004_1} Morozov O I 2004
  Symmetries of Differential Equations and Cartan's Equivalence  Method
   {\it Proc. Inst. Math. NAS of Ukraine}
   {\bf 50} Part 1, 196 - 203
\end{thebibliography}
\end{document}